\documentclass[aps,prb,twocolumn,groupedaddress]{revtex4}
\usepackage{epsfig}
\usepackage[american]{babel}
\usepackage{amsmath}

\newcommand{\Ha}{Hamiltonian}

\newcommand{\Hol}{Holstein}
\newcommand{\Hil}{Hilbert}
\newcommand{\Bri}{Brillouin}
\begin{document}

\title{Calculation of excited polaron states in the Holstein model}
\author{O. S. Bari\v si\' c}
\email{obarisic@ifs.hr}
\affiliation{Institute of Physics, Bijeni\v cka c. 46, HR-10000 Zagreb, Croatia}

\begin{abstract}

An exact diagonalization technique is used to investigate the low-lying excited
polaron states in the Holstein model for the infinite one-dimensional lattice.
For moderate values of the adiabatic ratio, a new and comprehensive picture,
involving three excited (coherent) polaron bands below the phonon threshold, is
obtained. The coherent contribution of the excited states to both the
single-electron spectral density and the optical conductivity is evaluated and,
due to the invariance of the Hamiltonian under the space inversion, the two are
shown to contain complementary information about the single-electron system at
zero temperature. The chosen method reveals the connection between the excited
bands and the renormalized local phonon excitations of the adiabatic theory, as
well as the regime of parameters for which the electron self-energy has notable
non-local contributions. Finally, it is shown that the hybridization of two
polaron states allows a simple description of the ground and first excited
state in the crossover regime.

\end{abstract}

\maketitle

\section{Introduction}\label{s04}

Although more than half of a century has elapsed since Landau proposed that the
charge carrier could be trapped by the distortion of a crystal
lattice,\cite{Landau} and Pekar introduced the term polaron,\cite{Pekar} a
number of questions regarding the single-polaron theory remain unanswered. This
is true even in the context of the \Hol\ \Ha,\cite{Holstein} one of the
simplest electron-lattice coupling models for the one-electron system. While the
literature pertaining to the ground state of such a model is extensive, much
less attention has been paid to the excited states, even at the low energies
for which they are most interesting.

The nature of the excited polaron states has been investigated within the
adiabatic approximation in Refs.\ \onlinecite{Kabanov,Kalosakas,Voulgarakis} by
neglecting the polaron translation. These works provide a simple picture of the
self-trapped polaron states for strong couplings. That is, the adiabatic
softening of the phonon modes within the self-trapped polaron states results in
several excitation energies below the bare phonon energy. It follows that the
lowest excitations of the system are the polaron states for which the electron
and the phonons are strongly correlated. When the polaron translation is
restored, each of the soft-phonon modes below the bare phonon energy can be
expected to develop the corresponding band if the local dynamics remain
adiabatic. Actually, moving polarons have been described within the adiabatic
approximation by neglecting the force impeding the polaron translational motion
due to the lattice discreteness.\cite{Melnikov,Shaw,Holstein3} However, the
band structure of the spectrum was not considered in these investigations. For
strong couplings, the bandwidth of the lowest band has been obtained by the
adiabatic theory in the context of the simplest two-site
model.\cite{Holstein2,Alexandrov}

When considering the band structure it is important to realize that the polaron
states are coherent in the range of energies below the phonon threshold (i.e.,
below the minimal energy for inelastic scattering).\cite{Engelsberg,Ciuchi} One
way to investigate their properties is by analyzing the single-electron and
optical conductivity spectra.  In the case of the single-polaron problem, the
low-frequency coherent part of the spectral weight directly determines the
polaron energies. Most of the previous spectral
investigations,\cite{Ranninger,Capone,Zhang,Wellein,Fehske,Hoffmann,Shawish}
however, make no mention of the excited coherent polaron bands. The exceptions
in this respect are provided by those works employing the dynamic mean-field
theory (DMFT), which is exact in the infinite dimension limit. The DMFT results
predict one excited polaron band below the phonon
threshold.\cite{Ciuchi,Fratini}

For the one-dimensional system, recent exact-diagonalization (ED) and
variational approaches\cite{Bonca,Barisic} obtained results for the lowest
state of the first excited band. Provided that the local electron dynamics
remain adiabatic, and that the adiabatic calculation of the self-trapped
polaron energies gives several solutions below the inelastic phonon threshold,
more than one excited polaron band is expected to occur in this energy range.
The present paper shows that the ED approaches for the infinite lattice, as
implemented in recent works,\cite{Wellein2,Bonca,Ku,Barisic} can be extended to
give very accurate results for the first few excited coherent polaron bands. In
the range of parameters for which the method converges, the coherent part of
the spectrum is found to exhibit up to three excited polaron bands, two more
than previously described. These results provide a better understanding of the
spatial and temporal structure of the polaron states, as well as of their
symmetry (parity) properties. The latter can be used to rationalize the
contributions of the bands to the low-energy single-electron and optical
conductivity spectra, respectively.

An additional and attractive feature of the excitation spectrum obtained herein
is its potential to help clarify the physical picture of the polaron crossover
for moderate values of the adiabatic ratio. It has been previously
proposed,\cite{Barisic} in the context of the variational
analysis,\cite{Proville,Barisic} that this crossover, in which the ground state
evolves from a light to a heavy polaron state, can be understood as the
anticrossing (hybridization) of two low-energy states. This question is
reconsidered here in terms of practically exact eigenstates and their
aforementioned parity properties.  Specifically, the hybridization is found to
occur between states of equal parity, whereas states of opposite parity cross
without any such hybridization. Furthermore, the hybridization of states in the
crossover regime is characteristic not only of the Holstein model under
consideration. Recently, for example, on the basis of accurate numerical
results obtained by the diagrammatic quantum Monte Carlo method\cite{Prokofev}
for the Rashba-Pekar polaron model,\cite{Pekar2} it has been suggested that the
polaron crossover in that instance also involves a hybridization of several
polaron solutions.\cite{Mishchenko}

ED methods, analogous to the one employed here, have found widespread
application in the ground-state calculations of numerous other many-body
problems. In this context, the present analysis of the excited states, together
with their symmetry attributes, is also of interest for the description of
low-energy eigenstates and spectral properties from a more general point of
view.

\section{\Hol\ polaron problem}\label{s06}

The one-dimensional \Hol\ model of interest here is defined by the \Ha

\begin{eqnarray}
\hat{H}&=&-t\sum_nc_{n}^{\dagger}\;(c_{n+1}+c_{n-1})+
\hbar\omega\sum_nb^{\dagger}_nb_n\nonumber\\&-&
g\sum_n c_n^{\dagger}c_n\;(b^{\dagger}_{n}+b_{n})
\label{Ham}\;.
\end{eqnarray}

\noindent $t$ is nearest-neighbor hopping energy of the electron, $\hbar\omega$
is the energy of dispersionless optical phonons, while $g$ is the
electron-phonon coupling energy. $c_n^{\dagger}$ and $b_n^{\dagger}$ are
creation operators for the electron and phonon at lattice site $n$,
respectively. As only the single-electron problem is treated, the spin indices
have been omitted.

The total momentum of the system ($\hat K$) is the sum of electron and phonon
momenta. As $\hat K$ commutes with the \Ha\ (\ref{Ham}), the complete set of
states of the system can be constructed from eigenstates of $\hat H$ and $\hat
K$. The \Ha\ (\ref{Ham}) is also invariant under space inversion. However, the
analysis of the resulting parity properties of the eigenstates will be 
deferred until later in the present discussion [Sec. \ref{s10d}].

The eigenstates that will be considered herein are those falling in the energy
window defined by

\begin{equation}
E^{(0)}_{K=0}\leq E<E^{(c)}\;,\;\;\;E^{(c)}=E^{(0)}_{K=0}+\hbar\omega\;.
\label{Ec}
\end{equation}

\noindent $E^{(0)}_{K=0}$ is the minimal energy of the system (the
zero-momentum polaron ground-state energy) and $\hbar\omega$ is the energy of
the bare phonon excitation. When the electron-phonon coupling is absent ($g=0$)
the states below the phonon threshold $E^{(c)}$ are those of the free electron
band. When $g$ is switched on, the free electron states evolve into those
states correlated with phonons. Furthermore, additional bands can appear below
$E^{(c)}$ as the coupling increases. All states in the energy window (\ref{Ec})
are correlated and will be referred to as polaron states. For $E\geq E^{(c)}$
the electron and the additional phonons can form weakly-bound states, which
results in a highly degenerate spectrum.

The energies of the coherent polaron states are henceforth denoted by
$E^{(i)}_K$, and the wave functions by $|\Psi_K^{(i)}\rangle$. $K$ is the
momentum of the polaron state (which is also the system momentum), while $i$
denotes the band index. The lowest (ground) polaron band will be denoted by
$i=0$. In the present paper the term ground state is used for the state of
minimal energy for a given momentum $K$, and not to the $K=0$ state only.
Accordingly, the first excited state corresponds to the first excited state of
a given momentum $K$, and so forth.

For the polaron bands below the phonon threshold $E^{(c)}$ the states
$|\Psi_K^{(i)}\rangle$ are the eigenstates of
$b_{q=0}$,\cite{Feinberg,Robin}

\[b_{q=0}|\Psi_K^{(i)}\rangle=\frac{1}{\sqrt N}\sum_nb_n|\Psi_K^{(i)}\rangle
=\frac{1}{\sqrt N}\frac{g}{\hbar\omega}|\Psi_K^{(i)}\rangle\;.\]

\noindent More generally, it can be shown that for all eigenstates $|\Psi_K^{E}\rangle$ the simple
sum-rule for the mean total lattice deformation, given by

\begin{equation}
\overline x_{tot}=\sum_n\overline x_n=x_0\sum_n
\langle\Psi_K^E|(b_n^\dagger+b_n)|\Psi_K^E\rangle=
\frac{2gx_0}{\hbar\omega}\;,\label{xtot}
\end{equation}

\noindent is satisfied. In Eq.\ (\ref{xtot}) $x_0$ is the space uncertainty of
the free harmonic oscillator with frequency $\omega$. Besides its physical
meaning, the sum-rule for the mean total lattice deformation can also be used
as a tool for checking the validity of results obtained with approximate
polaron wave functions.

\section{Exact translational method}\label{s08}

In the case of the \Hol\ polaron problem, the ED approach uses only a finite
number of states which contribute significantly to the exact polaron wave
function for a given set of \Ha\ parameters. By using the \Ha\ matrix
corresponding to the truncated (reduced) basis, and the appropriate numerical
scheme, one calculates the polaron wave functions and energies. The convergence
of the results can be verified by increasing the number of basis states in the
calculation. In most cases the results are very accurate, provided that the
truncation procedure is well chosen.

The ED method developed in the current paper is henceforth referred to as the
exact translational method ($eT$ method). The basis states of the $eT$ method
are given by\cite{Bonca}

\begin{eqnarray}
&&|n_0, n_{-1}, n_1,...,n_m\rangle_K=\nonumber\\
&&\;\frac{1}{\sqrt N}\sum_je^{iKja}c_j^\dagger|
n_0, n_{-1}, n_1,...,n_m\rangle_j\;.\label{TState}
\end{eqnarray}

\noindent The orthonormal wave function (\ref{TState}) describes an electron
which is surrounded by a cloud of phonons. The number of phonons at the $m$th
lattice site away from the electron is given by $n_m$, while $K$ corresponds to
the total system momentum and $a$ is the lattice constant. The basis states
(\ref{TState}) of different momenta $K$ are not mixed by the \Hol\ \Ha.

If the adiabatic ratio $t/\hbar\omega$ is not too large, the $eT$ method can be
used for studying polarons in the weak- and the strong-coupling regime, as well
as in the crossover regime between them. The present paper is focused,
particularly, on the regime in which $1\lesssim t/\hbar\omega\lesssim 5$ and
$g$ is arbitrary. For $g,\;t\gg\hbar\omega$, the number of relevant basis
states (\ref{TState}) becomes large and the problem of finding the polaron wave
functions becomes untreatable. Unfortunately, the limit $g,\;t\gg\hbar\omega$
which is not considered here, is difficult for other known numerical methods as
well.

The matrix representation of the \Hol\ \Ha\ in the $eT$ basis leads to a sparse
matrix. That is, it is straightforward to show that, by acting on the state
(\ref{TState}) with the \Ha\ (\ref{Ham}), the maximum number of  non-zero
matrix elements per $eT$ basis state is five. As an example, let us form the
reduced \Hil\ space of only five $eT$ states, i.e., of the zero-phonon state
$|0\rangle_K$, of three states with one phonon (phonon at the electron site, at
the left site from the electron, and at the right site from the electron),

\[|n_0=1\rangle_K\;, |n_{-1}=1\rangle_K\;, |n_1=1\rangle_K\;,\]

\noindent and finally, of the state with two phonons at the electron site
$|n_0=2\rangle_K$. The corresponding \Ha\ matrix is given by

\[\left [\begin{array}{ccccc}
-2t\cos{Ka}&g&0&0&0\\
g&\hbar\omega&-te^{-iKa}&-te^{iKa}&g\sqrt{2}\\
0&-te^{iKa}&\hbar\omega&0&0\\ 0&-te^{-iKa}&0&\hbar\omega&0\\
0&g\sqrt{2}&0&0&2\hbar\omega
\end{array}\right ]\;.\]

\noindent In the limit $g,\;t\ll\hbar\omega$ the $eT$ method gives a good
polaron ground state with only these five states. Nevertheless, for
$t\sim\hbar\omega$ one usually has to work with a truncated basis of quite
large dimension. In other words, the $eT$ method generally requires a numerical
scheme capable of dealing with large sparse matrices.

For this purpose the well-known Lanczos algorithm appears to be the most
appropriate choice.\cite{Cullum,Golub} Indeed, the previous papers using the
$eT$ method for ground-state calculations have employed this technique.  An
additional, attractive, feature of the Lanczos algorithm is that it is capable
of finding not just one, but rather a few extreme eigenvalues (and
eigenvectors) of sparse matrices, provided they are well separated (lying in
the discrete part of the spectrum). Accordingly, the current results have been
obtained by the Lanczos procedure with the so-called local orthogonalization.
In addition, the states have been calculated by the block-Lanczos procedure,
with both variants giving the same results.

It should be stressed that the present results are compared to the already
cited ground-state results\cite{Barisic,Bonca,Ku} of the $eT$ method. All
states are checked through the sum-rule (\ref{xtot}), as well as through their
mutual orthogonality.

\section{Polaron bands}\label{s10}

\subsection{Numerical results}\label{s10a}

\begin{figure*}[tbp!]
\begin{center}{\scalebox{0.62}{\includegraphics{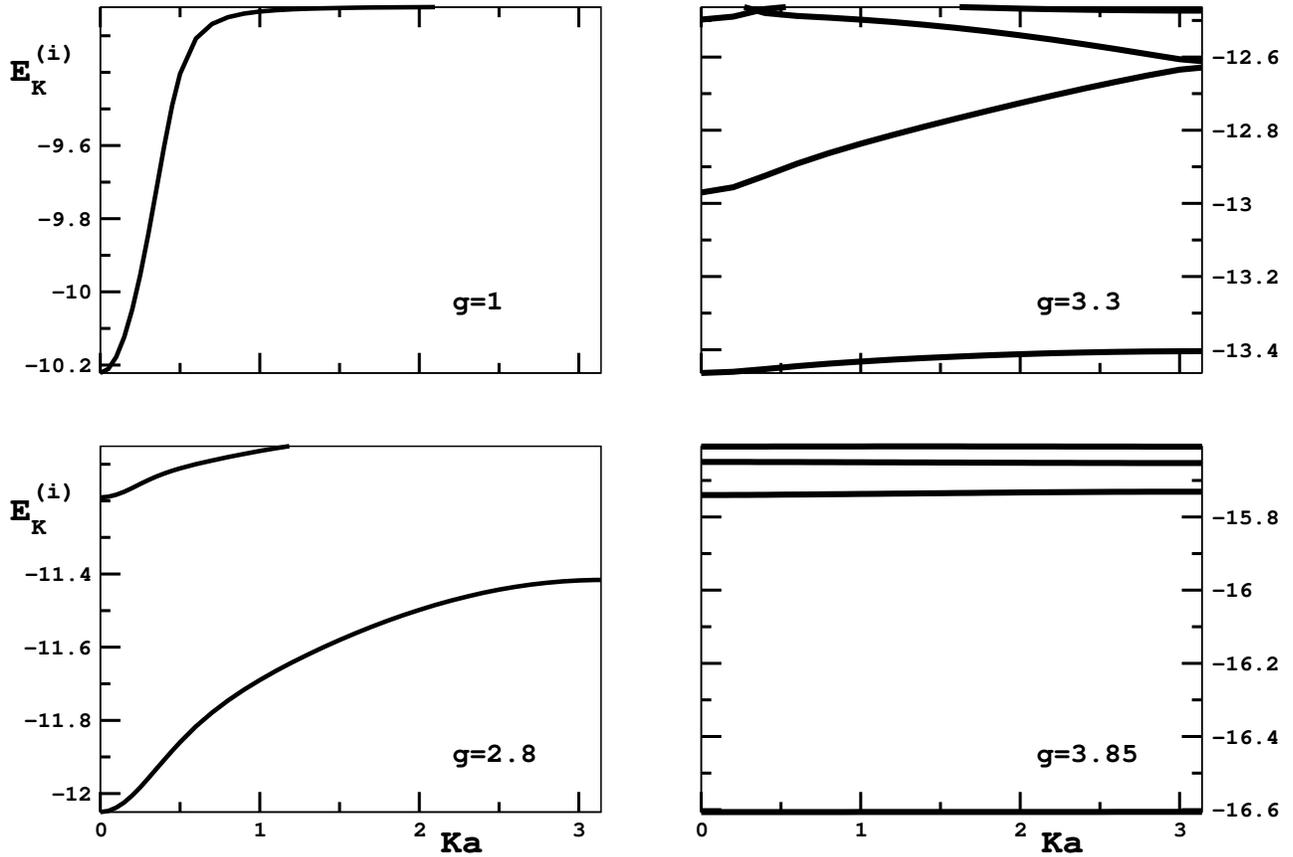}}}\end{center}

\caption{The polaron bands are plotted as functions of momentum ($K)$ for $t=5$
and four different values of $g$. $\hbar\omega=1$ defines the energy unit used
throughout this paper. The results for $g=1$ and $g=3.85$ correspond to the weak-
and the strong-coupling regimes, respectively, whereas results for $g=2.8$ and
$g=3.3$ correspond to the crossover regime. Only the part of spectrum below the
phonon threshold $E\leq E^{(c)}$ is shown.\label{fig1}}

\end{figure*}

At the beginning of our discussion it is instructive to observe how the polaron
bands are formed in regard to the strength of the electron-phonon coupling $g$.
For this purpose, the polaron bands are plotted in the four panels of Fig.\
\ref{fig1} as functions of the momentum ($K$), for $t=5$ and different values
of $g$. As in the remainder of the paper, $\hbar\omega=1$ is used as the energy
unit. Notice that, as discussed in connection with Eq.\ (\ref{Ec}), only the
part of the spectrum below the phonon threshold (between $E_{K=0}^{(0)}$ and
$E^{(c)}$) is shown.

In the weak-coupling limit for the lowest band ($i=0$), two regimes exist with
respect to the critical momentum $K_c$.\cite{Haken,Zhao} For $K\lesssim K_c$
the polaron state of energy $E_K^{(0)}<E^{(c)}$ is the ground state of the
system. For $K\gtrsim K_c$ the ground state consists of the polaron and the
unbound phonon excitation. The unbound phonon excitation carries the system
momentum $K$, while the polaron momentum is equal to $0$. For $K\gtrsim K_c$
the ground-state energy lies at the bottom of the incoherent part of the
spectrum $E\geq E^{(c)}$. The first panel of Fig.\ \ref{fig1} shows the $eT$
weak-coupling results. Notice that the top flat part of the lowest band for
$K\gtrsim K_c$ has been cut by the frame. The reason is that the numerical
error of the corresponding $eT$ states is slightly greater than that of the
zero-momentum ground state, while the energy interval shown is exactly equal to
the bare phonon energy $\hbar\omega$. As the electron-phonon coupling
increases, the lowest band becomes more renormalized, and finally, at some
critical coupling, the whole band falls bellow $E^{(c)}$.

By increasing electron-phonon coupling further, the additional polaron bands
emerge below $E^{(c)}$. In the second and the third panel of Fig.\ \ref{fig1}
the results for $g=2.8$ and $g=3.3$ are shown, respectively. These are the
choices of parameters that correspond to the crossover regime. For $g=2.8$,
only a part of the first excited band ($i=1$) lies below $E^{(c)}$, whereas for
$g=3.3$ there are four polaron bands in the relevant energy window. From the
third panel of Fig.\ \ref{fig1} ($g=3.3$) one sees that the top of the second
excited band ($i=2$) is at $K=0$, and the bottom is at $K=\pi/a$. In addition,
this band ($i=2$) crosses the other excited bands ($i=1,3$), i.e., the first
excited band near the end of the \Bri\ zone (for $g\lesssim 3.3$), and the
third excited band near the center of the \Bri\ zone.

For strong couplings ($g=3.85$) the $eT$ results are shown in the last panel of
Fig.\ \ref{fig1}. All the bands are very narrow.  Although it is hard to
distinguish the third excited band ($i=3$) from the plot-frame, note that
$E^{(3)}_K<E^{(c)}$ for all $K$.

\begin{figure}[bp]
\scalebox{0.3}{\includegraphics{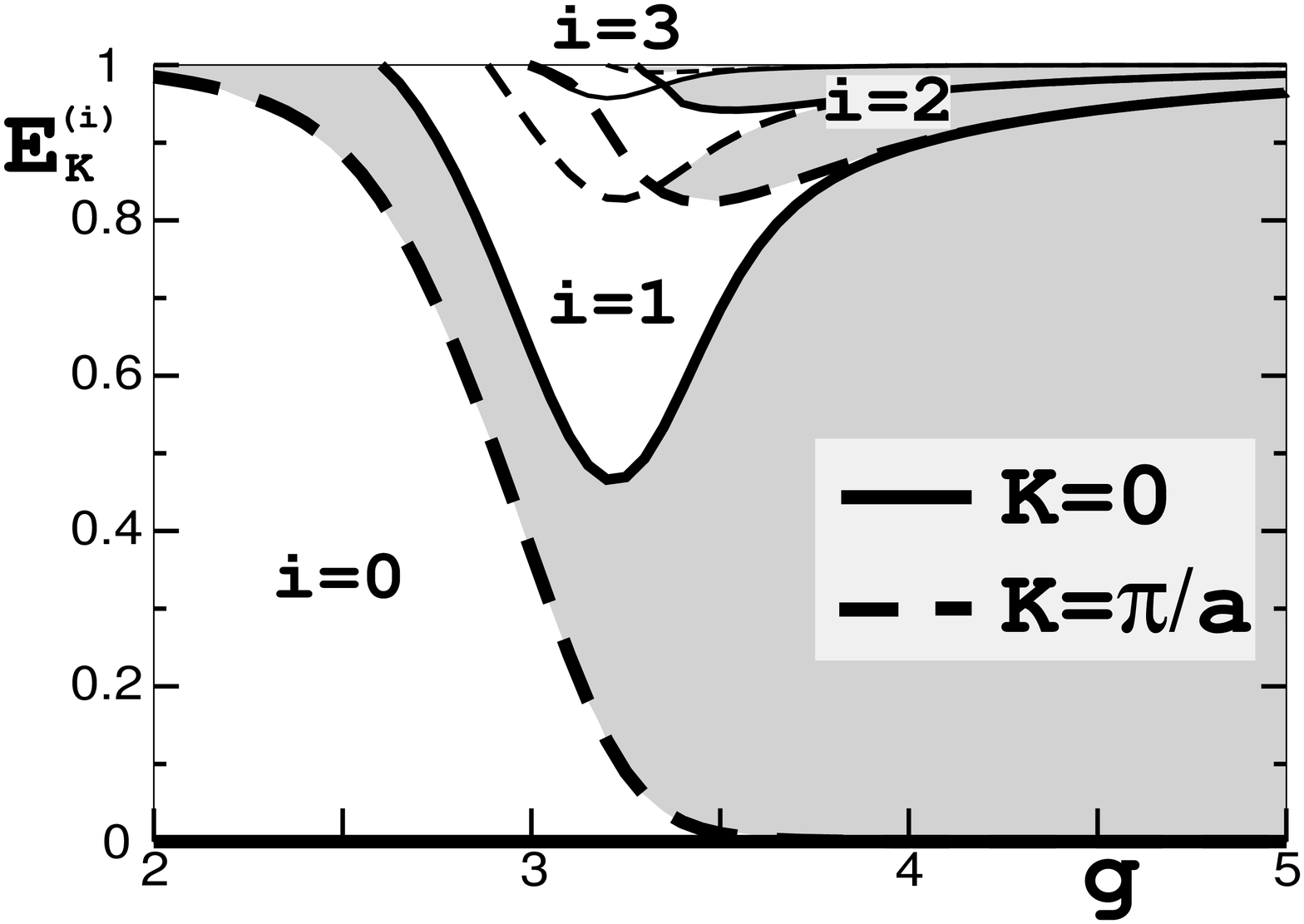}}

\caption{The polaron bands below the phonon threshold shifted by $E_{K=0}^{(0)}$
are plotted as functions of $g$ for $t=5$ ($\hbar\omega=1$). The band
boundaries correspond to $E_{K=0}^{(i)}$ and $E_{K=\pi/a}^{(i)}$ (except for
the highest excited band for which this is only approximately
true).\label{fig2}}

\end{figure}

Figure \ref{fig2} gives further insight into the polaron band formation. Here,
the $K=0$ and $K=\pi/a$ energy curves, shifted by $E_{K=0}^{(0)}$, are plotted
for four bands ($i$ is the band index). The results are presented for the
crossover and the strong-coupling regimes. In the weak-coupling regime (not
shown on Fig.\ \ref{fig2}) $E^{(0)}_{K=\pi/a}$ becomes smaller than $E^{(c)}$
for $g\approx 1.6$.

It can be seen from Fig.\ \ref{fig2} that the bandwidth of the lowest band
decreases continuously with $g$. On the other hand, the bandwidth of the first
excited state shows a more complicated behavior. Namely, for $g\approx 3.15$
this bandwidth is maximal, while, unlike for the lowest band, it decreases for
smaller values of $g$. Its maximal value and the minimum of
$E_{K=0}^{(1)}-E_{K=0}^{(0)}$ ($g\approx 3.2$) correspond to similar
electron-phonon couplings. From Fig.\ \ref{fig2} the band crossing which
involves the second and the two other excited bands can be clearly seen to
occur in the crossover regime. That fact that these bands cross each other
indicates that they belong to different symmetries, as will be discussed
further below [Sec. \ref{s10d}].

In strong-coupling regime the spectrum consists of very narrow bands. Although
in Fig.\ \ref{fig2} the bands are well separated, for very large couplings all
of the excited bands approach $E^{(c)}$ from below. Finally, it can be seen
from Fig.\ \ref{fig2} that, for $t=5$, the energy of the third excited band
stays close to $E^{(c)}$ both in the crossover and the strong-coupling regimes.
The $K$ values corresponding to the top of the third excited band depend
sensitively on the \Ha\ parameters. It should therefore be kept in mind that,
in contrast to the other bands, the $K=\pi/a$ curve only approximately
determines the boundary of this band. In particular, the third excited band is
slightly wider than suggested by Fig.~\ref{fig2} in the crossover regime.

\subsection{Strong coupling}\label{s10b}

In order to understand the physical background of the bands shown in Figs.
\ref{fig1} and \ref{fig2}, the strong-coupling and the crossover regime are
examined separately in the following sections. In the strong coupling regime
the time scale relevant to the polaron translation is much slower than the
time scale ($1/\omega$) involved in the local polaron dynamics. Therefore, the
local interplay between the electron density and the lattice deformation is
almost independent of the polaron momentum. From the energy point of view the
contributions corresponding to the polaron translation can be neglected, and
the polarons can be treated, to a good approximation, as self-trapped. The
deep potential well of the lattice deformation of the self-trapped polaron
captures the electron.  When the electron is light ($\hbar\omega< t$), it is
able to follow the slow motion of the lattice deformation (e.g., zero-point
motion), which results in the adiabatic renormalization of the phonon modes
within the lattice deformation.

The physical picture of the aforementioned self-trapped polaron follows from
the adiabatic theory, and can be considered to be well understood. Thus, in
order to achieve a better understanding of the current numerical results, it is
convenient to compare the renormalized phonon energies obtained in the
adiabatic limit to the spectrum calculated by the $eT$ method. For the \Hol\
polaron problem the renormalized phonon modes have been calculated by different
adiabatic
approximations.\cite{Melnikov,Shaw,Holstein3,Kabanov,Kalosakas,Voulgarakis,Alexandrov}
The procedure of Ref.\ \onlinecite{Kalosakas} (Born-Oppenheimer approximation
therein Sec. III) treats the lattice discreteness directly, while the polaron
translation is neglected. This approximation is appropriate for the present
case as the lattice discreteness is important for the small self-trapped
polarons, while the polaron translation has only a minor contribution to the
energy.

\begin{figure}[tbp]
\begin{center}{\scalebox{0.32}{\includegraphics{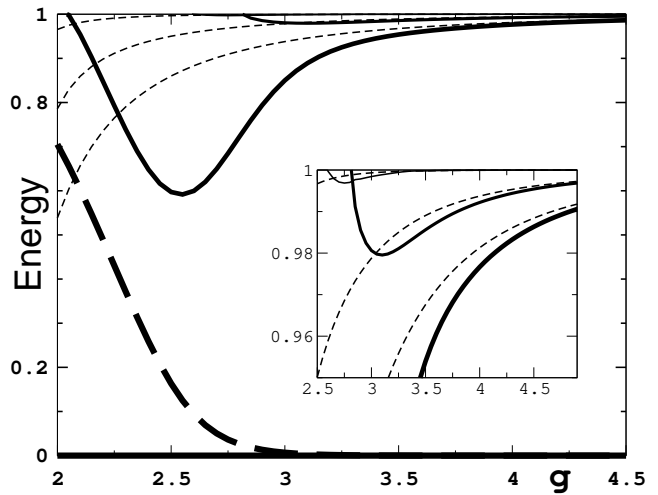}}}\end{center}

\caption{The energies of $K=0$ excited polaron states shifted by
$E_{K=0}^{(0)}$ (solid curves) are compared to the energies of the renormalized
phonon excitations obtained by the strong-coupling adiabatic approximation of
Ref.\ \onlinecite{Kalosakas} (short-dashed curves). The long-dashed curve
corresponds to $E^{(0)}_{K=\pi/a}$ and shows the lowest band narrowing. The
inset displays only a small part of the spectrum below $E^{(c)}=1$. $t=2.5$
($\hbar\omega=1$).\label{fig3}}

\end{figure}

The three short-dashed curves in Fig.\ \ref{fig3} are the renormalized phonon
energies of Ref.\ \onlinecite{Kalosakas}. The lowest excitation is a symmetric
vibration of the lattice with respect to the polaron center (breathing mode).
The next excitation is the antisymmetric vibration (pinning mode). The third
excited mode is again a symmetric vibration, although extended over a larger
number of lattice sites than in the case of the lowest excitation. In Fig.\
\ref{fig3}, the solid curves are the $eT$ energies of the $K=0$ polaron excited
states. In the strong coupling limit [the right part of Fig.\ \ref{fig3}] the
results of Ref.\ \onlinecite{Kalosakas} are recovered asymptotically, from
below. The inset of Fig.\ \ref{fig3}, where only a small part of the spectrum
below $E^{(c)}$ is shown, clearly demonstrates this behavior. For strong
couplings, the positions of the excited bands are given by the the symmetric
and antisymmetric phonon excitations of the adiabatic theory. In other words,
the local dynamics of the self-trapped polarons is adiabatic.

The energy $E_{K=\pi/a}^{(0)}$ is plotted as the long-dashed curve in Fig.\
\ref{fig3}. This curve defines the bandwidth of the lowest polaron band which
can be seen to become large in the crossover regime [the left part of Fig.\
\ref{fig3}], as the translation of the polaron becomes important. On the other
hand, in the strong coupling limit, the $eT$ method reproduces the narrow
cosine polaron bands,

\[E_{K}^{(i)}-E_{K=0}^{(i)}\approx 2t_{pol}^{(i)}\;[1-\cos{(Ka)}]\;,\]

\noindent where $t_{pol}^{(i)}$ is an effective polaron nearest-neighbor
hopping energy for the $i$th excited state.

\begin{figure}[tp]
\begin{center}{\scalebox{0.32}{\includegraphics{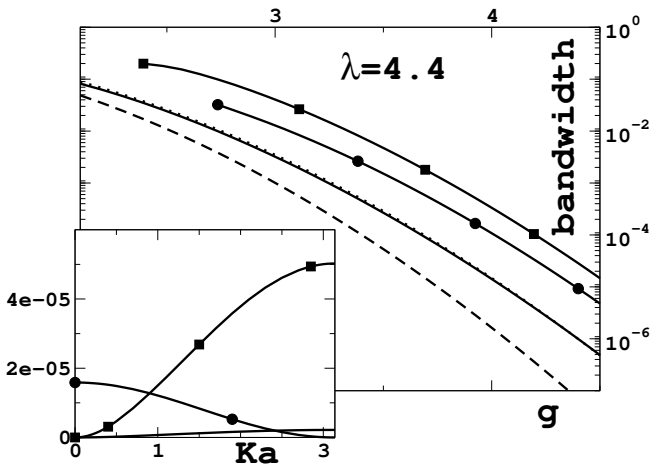}}}\end{center}

\caption{The bandwidths (solid curves) of the three lowest polaron bands are
plotted for $2\lesssim g\lesssim 5$ and $\lambda=g^2/t\;\hbar\omega=4.4$. Each
curve stops at the point where the top of the band crosses $E^{(c)}$. The
bandwidths of the first and the second excited band are denoted by squares and
circles, respectively. The long-dashed curve is the bandwidth of the
nonadiabatic small polaron [$\alpha=1$ in Eq.\ (\ref{tpol})]. In the inset, the
cosine-like ground and two excited bands of the self-trapped polaron are shown,
shifted by $E_{K=0}^{(i)}$, as functions of momentum $K$ ($g=4$, $t=5$,
$\lambda=3.2$). All the energies are in units of $\hbar\omega$.\label{fig4}}

\end{figure}

For strong couplings, the values of bandwidths may differ considerably from
band to band. In Fig.\ \ref{fig4} the bandwidths, as functions of $g$, are
compared for constant $\lambda=g^2/t\;\hbar\omega=4.4$. The solid curves are
the results for the three lowest polaron bands. One sees that the effective
hopping $t_{pol}^{(i)}$ is increased if the polaron (i.e., the local lattice
deformation) is excited. However, for strong couplings all bands are very
narrow ($4t_{pol}^{(i)}\ll\hbar\omega$), meaning that the polarons are very
heavy (self-trapped). In real solids the coherent transport of such band
states can be destroyed by the polaron-polaron interaction or
by imperfections.\cite{Alexandrov2,Mello2}

A close inspection of the $eT$ results in Fig.~\ref{fig4} reveals that the
hopping energy $t_{pol}^{(0)}$ of the self-trapped polaron for the lowest band
is given by the simple relation

\begin{equation}
t_{pol}^{(0)}=t\;\exp{[-(g/\hbar\omega)^2\;\alpha(\lambda)]}\;.\label{tpol}
\end{equation}

\noindent Equation (\ref{tpol}) is an extension of the well-known expression
for the nonadiabatic small-polaron hopping energy, which is obtained by setting
$\alpha=1$.\cite{Holstein2} $\alpha(\lambda)$ turns out to be a function of the
adiabatic parameter $\lambda$ only, i.e., $\alpha(\lambda)$ is an adiabatic
correction to the nonadiabatic small-polaron hopping. Notice that in the
adiabatic limit $\lambda$ defines the spreading of the lattice deformation (the
polaron size $\sim 1/\lambda$), as well as the renormalized phonon
energies.\cite{Kalosakas} For $\lambda=4.4$ the excellent fit to the $eT$
results in Fig. \ref{fig4} (the  short-dashed curve that follows almost exactly
the $eT$ results for the the lowest band) is achieved with $\alpha\approx
0.86$, even though the bandwidth of the lowest polaron band changes by several
orders of magnitude.

In summary, for the parameters under consideration [Fig.~\ref{fig4}], it
follows from Eq.\ (\ref{tpol}) that the translational dynamics of the lowest
band are essentially nonadiabatic, with adiabatic corrections. The narrow
bandwidth defines the slow time scale for the polaron hopping. In contrast,
Fig.~\ref{fig3} shows that the fast local dynamics of the self-trapped polarons
is adiabatic. The adiabatic softening of the local phonon modes determines the
positions of the ground and excited bands in the spectrum. The energy shift due
to the softening is much larger than the bandwidths, which means that the
polaron hopping can be neglected for the local dynamics.

\subsection{Crossover}\label{s10c}

In spite of the fact that the almost exact results for the lowest polaron band
in the crossover regime, obtained by various numerical methods, have been known
for quite some time, the corresponding qualitative explanation of the polaron
properties does not appear completely satisfactory. It has been suggested in
Ref.\ \onlinecite{Romero3}, by the use of global-local method,\cite{Brown} that
a simple empirical relation between \Ha\ parameters

\begin{equation}
g_{ST}\approx\hbar\omega+\sqrt{t\;\hbar\omega}\label{gST}
\end{equation}

\noindent describes the regime for which the fast change from light to heavy
polaron ground state takes place. Indeed, $g_{ST}$ predicts, to a good
approximation, the set of parameters for which the variation of the polaron
effective mass (as a function of $g$) is the fastest. 

For $t\gg\hbar\omega$ the crossover occurs in the adiabatic regime when
$g_{ST}\approx\sqrt{t\;\hbar\omega}$, i.e., $\lambda_{ST}\approx 1$.
Approaching $\lambda_{ST}$ from the strong-coupling adiabatic side
($\lambda>\lambda_{ST}$), the size of the polaron increases. Consequently, the
Peierls-Nabarro (PN) barrier decreases and the tunneling of the adiabatic
polaron to the neighboring sites becomes possible. This effect is also
responsible for the coupling of the pinning and the breathing lattice
modes.\cite{Barisic2} At $\lambda<\lambda_{ST}$, the restoring force of the
pinning mode due to the PN barrier can be neglected, and one may treat the
polaron as a freely moving.\cite{Melnikov,Shaw,Holstein3} Such a scenario thus
describes the crossover from the self-trapped (heavy) to propagating (light)
polaron states in the adiabatic limit. On the other hand, for
$t\ll\hbar\omega$, the nonadiabatic theory describes the polaron translation
for arbitrary $g$, and in particular, at $g=g_{ST}\approx\hbar\omega$ of Eq.\
(\ref{gST}). In this case, the polaron crossover appears essentially as a
passage from the weak- to the strong-coupling ground state.

A reinspection of Fig.\ \ref{fig2} shows that the two well-separated time
scales found in the strong-coupling regime, one related to the polaron
translation (given by the bandwidth), and the other related to the polaron
local dynamics (given by the renormalized phonon energies), become comparable
in the crossover regime. Furthermore, the behavior of the ground and excited
bands for $g\approx g_{ST}$ indicates that the polaron translation plays an
important role in the excitation spectrum. In other words, the influence of the
polaron hopping to the neighboring sites on the renormalized modes cannot be
neglected under these conditions, as it can for strong couplings. The
substantial difference in the crossover regime [see Fig.\ \ref{fig3}] between
the $eT$ results and the (adiabatically renormalized) phonon energies of Ref.\
\onlinecite{Kalosakas} is explained in this way. This difference, however, is
not necessarily related to nonadiabatic effects, as for the moderate values of
the adiabatic ratio under current consideration ($1\lesssim
t/\hbar\omega\lesssim 5$) it is not clear to what extent the nonadiabatic
dynamics enter into the description of the polaron translation.

\subsection{Parity}\label{s10d}

Irrespective of their temporal (or spatial) properties, when the polaron bands
cross and/or anticross as in Fig.\ \ref{fig2}, their symmetry properties become
important. As already mentioned, the \Hol\ \Ha\ is invariant under the space
inversion, and its eigenstates can be distinguished according their parity. A
simple linear transformation relates the momentum $|\Psi_K^{(i)}\rangle$ and
parity $|\Psi_K^{(i)}\rangle^P$ eigenstates,

\[|\Psi_K^{(i)}\rangle^P=|\Psi_K^{(i)}\rangle\pm\hat P|\Psi_K^{(i)}\rangle
=|\Psi_K^{(i)}\rangle\pm P\;|\Psi_{-K}^{(i)}\rangle\;,\]

\noindent so that

\begin{equation}
\hat P\;|\Psi_K^{(i)}\rangle=P\;|\Psi_{-K}^{(i)}\rangle\;.\label{Parity0}
\end{equation}

\noindent $\hat P$ denotes the space inversion operator, and $P=\pm 1$ are the
even- and odd-parity eigenvalues, respectively.

Considering the $eT$ method, the parity can be directly determined by
inspection of the wave function properties. The eigenstates
$|\Psi_K^{(i)}\rangle$ can be expanded in terms of the basis states
(\ref{TState}) with expansion coefficients $a_{n_0, n_{-1}, n_1,...,n_m}$,

\[|\Psi_K^{(i)}\rangle=\sum_{n_i}a_{n_0, n_{-1}, n_1,...,n_m}\;
|n_0, n_{-1}, n_1,...,n_m\rangle_K\;.\]

\noindent $\hat P$ acting on the basis state (\ref{TState}) gives

\[\hat P\;|n_0, n_{-1}, n_1,...,n_m\rangle_K=|n_0, n_1,
n_{-1},...,n_{-m}\rangle_{-K}\;.\]

\noindent Consequently, using Eq.\ (\ref{Parity0}), and the fact that
$|\Psi_{-K}^{(i)}\rangle$ is complex conjugate of $|\Psi_K^{(i)}\rangle$
(time-reversal), one finds that the expansion coefficients satisfy

\begin{equation}
a_{n_0, n_{-1}, n_1,...,n_m}=P\;a^*_{n_0, n_1,
n_{-1},...,n_{-m}}\;.\label{Parity}
\end{equation}

\noindent $a^*$ denotes the complex conjugate of $a$. The expansion
coefficients in Eq.\ (\ref{Parity}) stand for two local phonon configurations,
the first one is obtained from the second one when the phonons to the left and
right of the electron are interchanged ($n_m\rightarrow n_{-m}$).

Applying the above analysis within the framework of the $eT$ method yields
$P=-1$ for the second excited band ($i=2$), while $P=1$ for the ground and the
other two lowest excited bands ($i=0,1,3$). In this respect, the polaron bands
inherit the symmetry of the renormalized phonon modes as obtained in the
adiabatic limit. In Sec.~\ref{s10b} it was pointed out that the pinning mode
corresponding to the second excited band is an antisymmetric vibration of the
lattice (having odd parity), whereas the other two lowest excited modes are
symmetric vibrations (having even parity). Furthermore, one sees that the
crossing (rather than the anticrossing) between the excited bands in the
crossover regime, shown in Fig.~\ref{fig2}, involves bands of opposite symmetry
under the space inversion.

For $K=0$ and $K=\pi/a$ the linear transformation\cite{Hoffmann} of the basis
(\ref{TState})

\begin{eqnarray}
|n_0, n_{-1}, n_1,...,n_m\rangle_K^\pm&=&\frac{1}{\sqrt 2}\;(
|n_0, n_{-1}, n_1,...,n_m\rangle_K
\nonumber\\&\pm&|n_0, n_1, n_{-1},...,n_{-m}\rangle_K)\label{PMState}
\end{eqnarray}

\noindent defines two subspaces with different parities. It follows from Eq.\
(\ref{Parity}) that the $P=1$ eigenstates $|\Psi_{K=0,\pi/a}^{(i)}\rangle$
belong to the $+$ subspace of Eq.\ (\ref{PMState}), while the $P=-1$
eigenstates belong to the $-$ subspace. One sees that for $K=0$ and $K=\pi/a$
the parity actually defines the symmetry of the local phonon configuration with
respect to the electron. For $P=1$ this configuration is symmetric, while it is
antisymmetric for $P=-1$.

\subsection{Anticrossing}\label{s10e}

In Ref.\ \onlinecite{Barisic} the correlated behavior of the ground and the
first excited state in the crossover regime was analyzed for moderate values of
the adiabatic ratio using a method based on the variational approach. It was
argued that the anticrossing of two physically different polaron states, one
(heavy) for which the translation energy is almost negligible, and the other
(light) for which this energy is important, can describe in simple terms the
mechanism of the crossover. It follows from this interpretation that by
increasing $g$ in the critical region of parameters near $g_{ST}$ [Eq.\
(\ref{gST})], the contribution to the ground state from the light state
decreases in favor of the heavy state which has lower energy at stronger
couplings. On the other hand, the opposite change occurs for the first excited
state, which is heavier than the ground state for $g\lesssim g_{ST}$, while
being lighter for $g\gtrsim g_{ST}$.

\begin{figure}[tbp]
\begin{center}{\scalebox{0.32}{\includegraphics{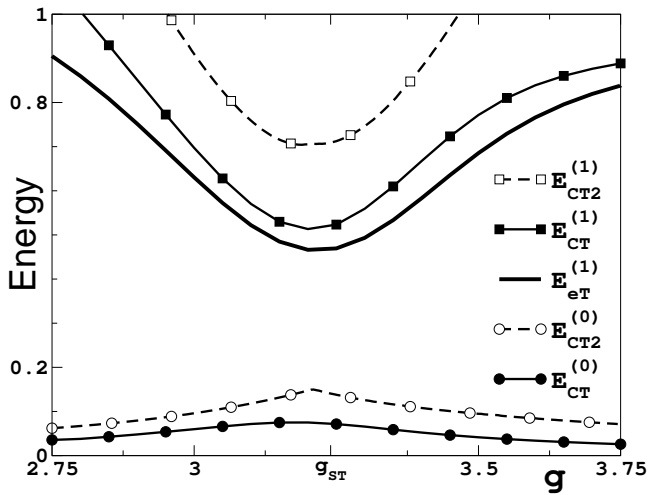}}}
\end{center}

\caption{The ground- and first excited-state energies obtained by the $CT$
method of Ref.\ \onlinecite{Barisic}, and the first excited-state energy
obtained by the $eT$ method, plotted as functions of $g$. The chosen set of
parameters is the same as for Fig. 3 of Ref.\ \onlinecite{Barisic}, $t=5$,
$g_{ST}\approx3.24$ ($\hbar\omega=1$), $K=0$. All curves are shifted by the
ground-state $eT$ energy.\label{fig5}}

\end{figure}

In Fig.\ \ref{fig5} the polaron ground and first excited energies obtained in
Ref.\ \onlinecite{Barisic}, denoted by $CT$, are compared to those of the $eT$
method. Although the $CT$ energies may be considered as a fair approximation,
the anticrossing picture can be discussed more accurately in the context of the
current $eT$ results. One sees that the $eT$ results confirm that the minimum
of $E_{K=0}^{(1)}-E_{K=0}^{(0)}$, the two states of the same symmetry ($P=1$),
corresponds to $g\approx g_{ST}$. It is important to notice that the $P=-1$
states of the second excited band ($i=2$) [Fig. \ref{fig2}] are not involved in
the anticrossing. The rest of the investigation presented here is focussed on
polaron effective mass and on the properties of the lattice deformation.

\begin{figure}[tp]
\begin{center}{\scalebox{0.32}{\includegraphics{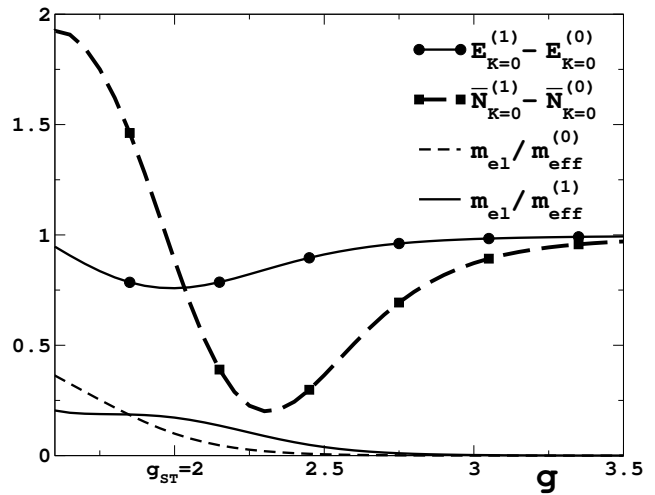}}}\end{center}

\caption{Difference of the energy, the mean number of phonons (\ref{Ntot}), and
the effective mass (\ref{meff}) of the ground and the first excited state for
$t=1$ ($\hbar\omega=1$). The minimum of $E_{K=0}^{(1)}-E_{K=0}^{(0)}$ and
$\overline{N}_{K=0}^{(1)}=\overline{N}_{K=0}^{(0)}+1$ almost coincide with
$g_{ST}=2$. Notice that for $g\lesssim g_{ST}$ the lowest excited state becomes
heavier than the ground state ($m_{eff}^{(1)}>m_{eff}^{(0)}$).\label{fig6}}

\end{figure}

The polaron effective mass $m_{eff}^{(i)}$ can be evaluated numerically by
using the relation\cite{Fehske,Bonca}

\begin{equation}
m_{el}/m_{eff}^{(i)}=\frac{1}{t}\;\frac{E^{(i)}_{K=\Delta
k}-E^{(i)}_{K=0}}{(\Delta k\;a)^2}\;,\label{meff}
\end{equation}

\noindent where $\Delta k$ is a small finite deviation of the momentum from
the value $K=0$. $m_{el}$ is the effective mass of the electron, and $i$
denotes the band number. The ground- ($m_{eff}^{(0)}$) and the first
excited-state ($m_{eff}^{(1)}$) effective masses are compared for $t=1$ in
Fig.\ \ref{fig6}. Although $m_{eff}^{(0)}>m_{eff}^{(1)}$ for $g\gtrsim
g_{ST}$, $m_{eff}^{(1)}$ becomes smaller than $m_{eff}^{(0)}$ as $g$
decreases. For larger $t$, the results are qualitatively the same [e.g., see
Fig.\ \ref{fig9}], which means that the anticrossing picture indeed matches
the behavior of $m_{eff}^{(0)}$ and $m_{eff}^{(1)}$ near $g_{ST}$.

For fixed $g$, the mean total lattice deformation (\ref{xtot}) is the same for
the ground and for the excited states. However, the local phonon cloud around
the electron can be more or less localized, which affects the polaron hopping
to the neighboring sites. A light state implies that the associated local
lattice deformation is spread to a larger number of lattice sites than for the
heavy state. Namely, such a deformation gives rise to a greater effective
hopping integral.

The mean number of phonons in the polaron state is given by

\begin{equation}
\overline{N}_{K}^{(i)}=\langle\Psi_K^{(i)}|
\sum_mb^{\dagger}_mb_m|\Psi_K^{(i)}\rangle\;.\label{Ntot}
\end{equation}

\noindent In the adiabatic limit, this number is quadratic in the local lattice
deformation for the ground state. A more localized (heavier) lattice
deformation leads to larger $\overline{N}_{K}^{(0)}$. Within the adiabatic
approximation the excited renormalized phonon should, in general, increase the
mean number of phonons with respect to $\overline{N}_{K}^{(0)}$. Particularly,
in the strong-coupling limit one obtains

\begin{equation}
\lim_{g\rightarrow\infty}\overline{N}_{K}^{(1)}=\overline{N}_{K}^{(0)}+1\;,
\label{Np1}
\end{equation}

\noindent as the renormalization of the lattice vibrations becomes negligible.

\begin{figure}[tbp]
\begin{center}{\scalebox{0.31}{\includegraphics{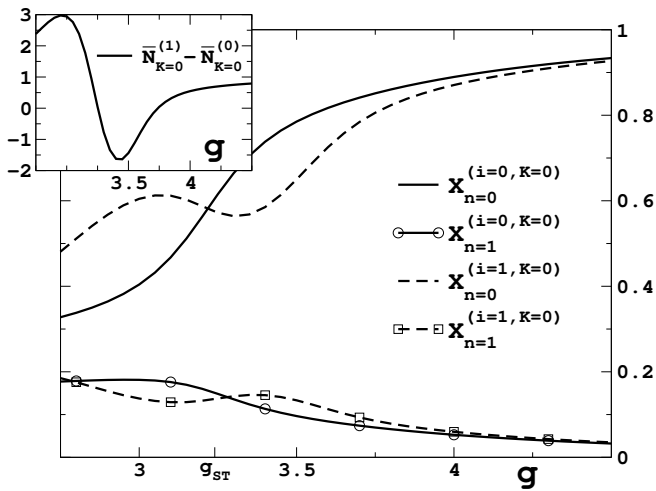}}}\end{center}

\caption{The electron-lattice deformation correlation function $X^{(i,K)}_n$,
Eq. (\ref{Xn}), plotted as a function of $g$ for $n=0$ and $n=1$. In the former
case $X^{(i,K)}_n$ gives the correlation at the electron site, while in the
latter it gives the correlation between the electron and the lattice
deformation at the neighboring site. The difference between the mean number of
phonons in the ground and the first excited state is shown in the inset. $t=5$,
$K=0$, $g_{ST}\approx 3.24$ ($\hbar\omega=1$).\label{fig7}}
\end{figure}

From Fig.\ \ref{fig6}, in which $\overline{N}_{K}^{(1)}-\overline{N}_{K}^{(0)}$
is plotted as a function of $g$, one sees that the $eT$ results tend to Eq.\
(\ref{Np1}) for strong couplings. In the crossover regime, on the other hand,
$\overline{N}_{K}^{(1)}-\overline{N}_{K}^{(0)}$ deviates from Eq.\ (\ref{Np1})
considerably. $\overline{N}_{K}^{(1)}$ is greater than
$\overline{N}_{K}^{(0)}+1$ for $g\lesssim g_{ST}$ and smaller than
$\overline{N}_{K}^{(0)}+1$ for $g\gtrsim g_{ST}$. For larger $t$, the amplitude
of the deviation increases even further. As shown in in the inset of Fig.\
\ref{fig7}, $\overline{N}_{K}^{(1)}<\overline{N}_{K}^{(0)}$ for $3.25\lesssim
g\lesssim 3.75$, while the minimal and maximal values of
$\overline{N}_{K}^{(1)}-\overline{N}_{K}^{(0)}$ define an interval of almost 5
phonons.

Additional insights into the local polaron properties can be provided by
studying the appropriate electron-lattice correlation function,

\begin{equation}
X^{(i,K)}_n=\frac{\hbar\omega}{2g}\langle\Psi_K^{(i)}|
\sum_mc^\dagger_mc_m\;(b^{\dagger}_{m+n}+b_{m+n})|\Psi_K^{(i)}\rangle\;.\label{Xn}
\end{equation}

\noindent $X^{(i,K)}_n$ is the mean lattice deformation induced at the $n$th
site away from the electron. The correlation function (\ref{Xn}) is normalized
in such a way that $\sum_nX^{(i,K)}_n=1$. As the correlation spreads to
adjacent lattice sites, the correlation at the electron site $X^{(i,K)}_{0}$
decreases. Consequently, $X^{(i,K)}_{0}$ should be larger for the heavy than
for the light polaron state.

The ground- and first excited-state results for $X^{(i,K=0)}_n$ are plotted in
Fig.\ \ref{fig7}. The relationship between $X^{(0,0)}_{0}$ and $X^{(1,0)}_{0}$
changes near $g_{ST}$. This is exactly what one would expect from the
anticrossing picture. $X^{(0,0)}_{0}<X^{(1,0)}_{0}$ for $g\lesssim g_{ST}$. As
the contribution of the heavy state to the ground state becomes dominant for
$g\gtrsim g_{ST}$, the relationship changes, and
$X^{(0,0)}_{0}>X^{(1,0)}_{0}$. The spreading of the lattice deformation as a
function of $g$ may also be deduced from $X^{(0,0)}_{1}$ and $X^{(1,0)}_{1}$.
The results corresponding to this lattice site, shown in Fig.\ \ref{fig7},
again lead one to the same conclusion. For $g\lesssim g_{ST}$ the lattice
deformation seems to spread more for the ground than for the excited state,
and vice-versa for $g\gtrsim g_{ST}$.

All of the aforementioned ground- and first excited-state properties (the
energy, the effective mass, the mean number of phonons, and the
electron-lattice deformation correlation function) indicate that for
$t/\hbar\omega\lesssim 5$ the anticrossing of two (light and heavy) polaron
states occurs near $g_{ST}$. Although criteria such as
$E^{(1)}_{K=0}-E^{(0)}_{K=0}$ being minimal, $m_{eff}^{(1)}=m_{eff}^{(0)}$,
$X^{(0,0)}_{0}=X^{(1,0)}_{0}$, or alternative criteria, do not agree exactly,
they all predict the polaron crossover to occur within the same very narrow
parameter range, given approximately by Eq.\ (\ref{gST}).

\section{Electron properties}\label{s15}

\subsection{Single-electron spectral function}\label{s15a}

\begin{figure}[bp]
\begin{center}{\scalebox{0.31}{\includegraphics{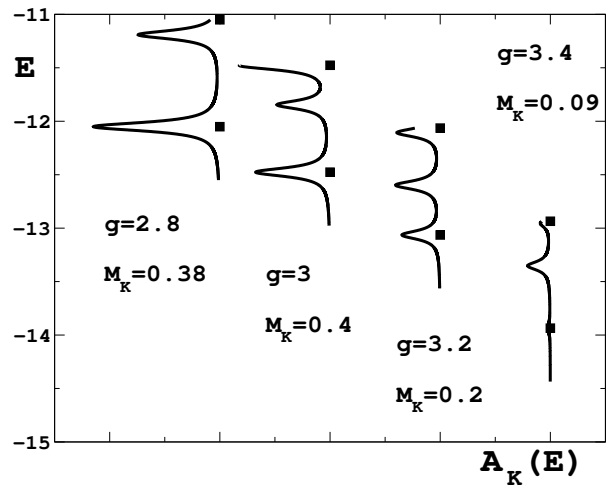}}}\end{center}

\caption{$A_K^<(E)$ given by Eq. (\ref{AK}) is plotted for $t=5$, $K=0$, and 4
values of $g$ in the crossover regime. All spectra are broadened by a
Lorentzian of width 0.05. For each curve the lower square denotes
$E=E^{(0)}_{K=0}$, while the higher one denotes $E=E^{(c)}$. Notice that for
$g\gtrsim g_{ST}\approx 3.24$, the spectral weight associated with the first
excited state is larger than the one associated with the ground state. $M_K$,
given by Eq.\ (\ref{MK}), is the total normalized spectral weight of the ground
and excited polaron states (below the phonon threshold).\label{fig8}}

\end{figure}

Denoting the vacuum state by $|0\rangle$, the zero-temperature single-electron
Green function can be expressed as\cite{Alexandrov1,Kornilovitch}

\[G_K(E)=\langle 0|c_K\frac{1}{E-\hat H+i0^+}c^{\dagger}_K|0\rangle\;.\]

\noindent For the energies below the phonon threshold $E<E^{(c)}$ no
polaron-phonon scattering takes place, which has the consequence that the
imaginary part of the electron self-energy $\Sigma_K(E<E^{(c)})$ tends to
zero.\cite{Engelsberg,Ciuchi} Accordingly, the low-energy part of the spectral
function $A_K(E)$, denoted by $A_K^<(E)$, is defined by the simple poles of
$G_K(E)$ at $E=E_K^{(i)}$,

\begin{equation}
A_K^<(E<E^{(c)})=\sum_i|\langle\Psi_K^{(i)}|c^{\dagger}_K\rangle|^2\;
\delta(E-E_K^{(i)})\;,\label{AK}
\end{equation}

\noindent i.e., the spectral function $A_K^<(E)$ is defined by the polaron
energy $E_K^{(i)}$ and the quasiparticle weight
$Z_K^{(i)}=|\langle\Psi_K^{(i)}|c^{\dagger}_K\rangle|^2$. In particular, for
$K=0$ and $K=\pi/a$, the quasiparticle weight $Z_{K=0,\pi/a}^{(i)}$ vanishes by
parity for the $P=-1$ eigenstates $|\Psi_{K=0,\pi/a}^{(i)}\rangle$, because the
free electron (zero-phonon) state $|c^{\dagger}_{K=0,\pi/a}\rangle$ belongs to
the $+$ subspace of Eq.\ (\ref{PMState}). 

In Fig.\ \ref{fig8} $A_{K=0}^<(E)$ is plotted as a function of $E$ for a few
different values of $g$ in the crossover regime. The spectral weight
corresponding to the ground polaron state is of the same order of magnitude as
the one corresponding to the excited (even-parity) polaron states. Moreover,
$M_K$ defined by

\begin{equation}
M_K=\int A_K^<(E)\;dE\;,\label{MK}
\end{equation}

\noindent reveals that, in the crossover regime (for $g\lesssim g_{ST}$),
nearly 40 percent of the total spectral weight confined to the ground and
excited polaron states can be located in the energy window below the phonon
threshold. Therefore, at least for $g\sim g_{ST}$ and for the values of
adiabatic ratio under current considerations ($t/\hbar\omega\lesssim 5$), the
results imply the existence of a few well-pronounced peaks in the
single-electron spectral density below the phonon threshold. For stronger
couplings, on the other hand, the quasiparticle weight below the phonon
threshold is almost completely suppressed.

\subsection{Electron self-energy}\label{s15b}

When the electron propagator $G_K(E)$ is expressed in terms of the electron
self-energy $\Sigma_K(E)$, both $Z_K^{(i)}$ and $E_K^{(i)}$ of Eq.\ (\ref{AK})
can be related to $\Sigma_K(E)$. E.g., the polaron effective mass
$m_{eff}^{(i)}$ is given in terms of the self-energy $\Sigma_K(E)$ by the
standard formula\cite{Mahan}

\begin{equation}
m_{el}/m_{eff}^{(i)}
=\frac{1+\partial_{\varepsilon_K}\Sigma_K(E)}
{1-\partial_E\Sigma_K(E)}=Z_{K=0}^{(i)}\;(1+\gamma^{(i)})\;,\label{meff2}
\end{equation}

\noindent where $1/Z_K^{(i)}=1-\partial_E\Sigma_K(E)$, and
$\varepsilon_K=K^2a^2t=\hbar^2K^2/2m_{el}$. In Eq. (\ref{meff2}) $i$ is used to
distinguish between results for the ground and excited (even-parity) states.
$\gamma^{(i)}$ is the abbreviation for the appropriately normalized
$K$-dependent contribution to $m_{eff}^{(i)}$.

\begin{figure}[tbp]
\begin{center}{\scalebox{0.32}{\includegraphics{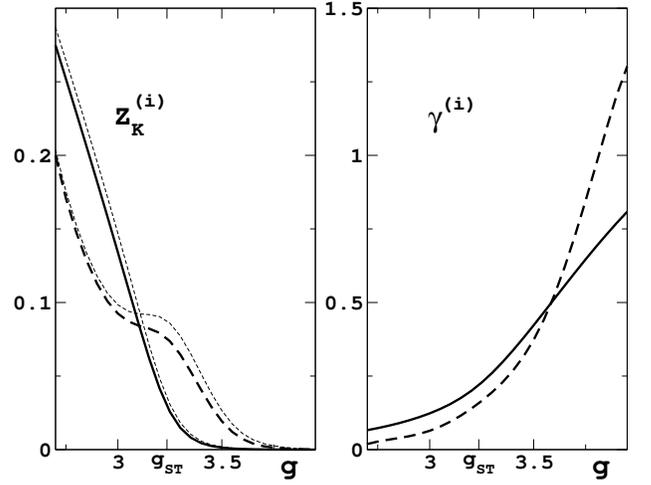}}}\end{center}

\caption{$Z_{K=0}^{(i)}$ and $\gamma^{(i)}$ given by Eq.\ (\ref{meff2})
are plotted in the left and right panel, respectively ($t=5$, $g_{ST}\approx
3.24$, $\hbar\omega$=1). The solid curves are the ground-state results, the
long-dashed are the first excited-state results. In the left panel the
short-dashed curves show the inverse effective mass
$m_{el}/m_{eff}^{(i)}$.\label{fig9}}

\end{figure}

When the self-energy $\Sigma_K(E)$ is local ($K$-independent, $\gamma^{(i)}=0$)
the polaron effective mass is related solely to the quasiparticle weight
$Z_{K=0}^{(i)}$. Thus, the non-local character of the self-energy can be
investigated by comparing $\gamma^{(i)}$ to unity.\cite{Ku} Such analysis is
particularly interesting in the context of DMFT, as the locality (implying
$\gamma^{(i)}=0$) of the self-energy is an essential ingredient of this
approach.\cite{Ciuchi} In both, the weak-coupling ($g\rightarrow 0$) and
nonadiabatic ($t\rightarrow 0$) limits, $\Sigma_K(E)$ becomes
local.\cite{Ciuchi,Fehske,Alexandrov1} Within the Holstein-Lang-Firsov
approximation\cite{Holstein2,Alexandrov1} and for $t\rightarrow 0$, the
self-energy $\Sigma_K(E)=\Sigma(E)$ is, to a good approximation, given by its
expansion to the first order in $E$ over the whole energy range defined by the
lowest band. In this case, $Z_{K}^{(0)}\approx Z^{(0)}$ defines the
renormalization of the narrow cosine-like lowest band for any $K$.

With the $eT$ method $\gamma^{(i)}$ can be obtained in terms of $Z_{K=0}^{(i)}$
and $m_{eff}^{(i)}$, the latter being estimated numerically from the band
dispersion at the center of the \Bri\ zone [Eq.\ (\ref{meff})]. The $eT$
results agree with the aforementioned analytical findings for small $t$ and/or
$g$.\cite{Ku} It follows that significant non-local contributions to
$\Sigma_K(E)$ can possibly occur in the regime where neither $t$ nor $g$ are
small. In the left panel of Fig.\ \ref{fig9} $Z_{K=0}^{(i)}$ is plotted for
moderate $t$ as function of $g$, while the right panel shows $\gamma^{(i)}$ for
the same set of parameters. Although $\gamma^{(i)}$ does not contribute to
$m_{eff}^{(i)}$ substantially for $g\lesssim g_{ST}$, at stronger couplings
$\gamma^{(i)}$ reaches values of the order of unity for both the ground and the
first excited state. That is, the non-local contribution to the electron
self-energy is as equally important as the local one. However, it should be
noticed that, in this regime, the quasiparticle weight $Z_{K}^{(i)}$ becomes
small.

\subsection{Optical conductivity}\label{s15c}

The interband optical conductivity may be evaluated within the linear-response
theory from the current-current correlation function. The low-energy (coherent)
contribution to the real part of the interband conductivity at zero temperature
is given by\cite{Maldague,Zhang}

\begin{equation}
\sigma_R^<(E)=\sum_{i\neq 0}\frac{|\langle\Psi^{(i)}|\hat
J|\Psi^{(0)}\rangle|^2}{E^{(i)}-E^{(0)}}\;\delta(E-E^{(i)}+E^{(0)})\;,\label{sK}
\end{equation}

\noindent where $|\Psi^{(i)}\rangle$ are the $K=0$ polaron states, herein
calculated by the $eT$ method. The incoherent contribution to the interband
conductivity, denoted by $\sigma_R^>(E)$, corresponds to the continuum of the
excitation spectrum above the phonon threshold. Since the current operator in
Eq.\ (\ref{sK}), defined by

\[\hat J=it\sum_n (c^\dagger_{n+1}c_n-c^\dagger_nc_{n+1})\;,\]

\noindent is odd under the space inversion, the nonvanishing matrix elements in
Eq.\ (\ref{sK}) are those between the $P=1$ ground state $|\Psi^{(0)}\rangle$
and $P=-1$ excited states $|\Psi^{(i)}\rangle$. Actually, the incoherent part
of the interband optical conductivity $\sigma_R^>(E)$ involves only $P=-1$
excitations as well. On the contrary, as argued in Eq.\ (\ref{AK}), the
spectral function $A_K(E)$ contains information about the $P=1$ part of the
spectrum at $K\approx 0$. The optical conductivity at zero temperature is
therefore described by those excited states for which the quasiparticle weight
$Z_{K\approx 0}$ vanishes, i.e., those states that are not seen in the
single-electron spectral function.

The real part of the optical conductivity $\sigma_R(E)$ includes the interband
part and the (intraband) Drude term at $E\approx 0$. It follows from the
well-known sum-rule\cite{Maldague} that the total spectral weight of
$\sigma_R(E)$ is given by the mean value of the electron kinetic energy in the
ground state. Consequently, not only is the $eT$ method capable of calculating
$\sigma_R^<(E)$, but the total spectral weight of $\sigma_R(E)$ is also
accessible. Furthermore, one may estimate the weight of the Drude term from the
polaron effective mass.\cite{Shawish}

According to the $eT$ results, for $t\leq 5$ there is only one interband
transition $|\Psi_{K=0}^{(0)}\rangle^+\rightarrow |\Psi_{K=0}^{(2)}\rangle^-$
which contributes at zero temperature to $\sigma_R^<(E)$ [Eq. (\ref{sK})]. It
is interesting to find the corresponding non-Drude spectral weight below the
phonon threshold. However, $E_{K=0}^{(2)}$ is less than $E^{(c)}$ for
relatively strong couplings, for which most of the spectral weight of
$\sigma_R(E)$ belongs to high energies. In this parameter regime only a few
percent (or less) of the total spectral weight of $\sigma_R(E)$ corresponds to
$\sigma_R^<(E)$, or to the Drude term. This simple low-energy picture of
$\sigma_R(E)$ becomes certainly more complex at finite temperature,
particularly in the crossover regime where the bandwidths of the coherent
polaron states increase. Studies at finite temperature, however, require
different methods than the one presented here.

\section{Summary}\label{s16}

This paper reports an exact-diagonalization study of the ground and excited
polaron states in the one-dimensional \Hol\ model.  For values of the adiabatic
ratio $t/\hbar\omega\lesssim 5$, accurate energies and wave functions are
obtained for translationally invariant solutions of the infinite lattice
problem. The chosen method is restricted to the part of the spectrum below the
phonon threshold, for which there are no phonon excitations uncorrelated to the
polaron.

The eigenstates of the \Hol\ \Ha\ can be distinguished according to their
parity. This property follows form the \Ha\ which is invariant under the space
inversion. It is shown that only the odd excited states of the system are
relevant to the optical conductivity at zero temperature. However, the
contribution of the coherent excited polaron state is rather small with respect
to the total spectral weight. On the other hand, it is the even states that
contribute to the single-electron spectral function for $K\approx 0$. In this
case, the contribution of the coherent excited polaron states is found to be
important, particularly in the crossover regime.

In the strong-coupling regime the results agree with the picture of
self-trapped polarons, which may be regarded as well understood. The spectrum
is characterized by very small polaron bandwidths. Consequently, the time scale
associated with the polaron translation is very large, and the contributions to
the fast local polaron dynamics from the polaron hopping are negligible. By
making a comparison with the $eT$ results, it is shown that the adiabatic
theory (the Born-Oppenheimer approximation) provides a good description of the
local properties of the self-trapped polaron. Namely, the energies of the
renormalized phonon excitations are close to the $eT$ excited-band energies.
Two of the excited bands correspond to the symmetric adiabatic vibrations of
the lattice, while one corresponds to the antisymmetric (pinning) vibration.
Finally, it is found that for strong couplings the electron self-energy shows
significant non-local ($K$-dependent) behavior at moderate $t$.

As the electron-phonon coupling decreases at moderate $t$, the simple band
structure found in the strong-coupling regime (i.e., the narrow and well
separated polaron bands) evolves notably. The effective hopping integral of the
ground and excited polaron states defines a time scale which, in the crossover
regime, is comparable to the time scale relevant for the local polaron
dynamics.  In the crossover regime the excited (odd-parity) band which
corresponds to the pinning vibration in the strong-coupling regime, crosses the
other two (even-parity) bands which correspond to symmetric vibrations. The
results suggest that the same critical set of parameters defined by $g_{ST}$
[Eq.\ (\ref{gST})], found for the polaron ground-state crossover, may also be
associated with the rapid change of the low-energy excitation spectrum.

Aside from the fact that the polaron translation and the local dynamics mix in
the crossover regime, the intermediate (moderate) values of the adiabatic ratio
employed herein present additional difficulties for qualitative understanding.
Namely, for $1\lesssim t/\hbar\omega\lesssim 5$, both the adiabatic and the
nonadiabatic contributions could be important for the polaron crossover.
Nevertheless, in the context of the $eT$ approach, the analyses of the energy,
the effective mass, and the lattice deformation properties, all agree in one
important respect. That is, for moderate values of the adiabatic ratio, two
polaron states of even parity (one heavy and one light) undergo an anticrossing
in the same region of parameters for which the ground state crosses over from
the weak- to the strong-coupling regime.

\begin{acknowledgments}
\label{ack}

This work was supported in part by the project SCOPES-7KPJ065619 of the Swiss
National Science Foundation. The calculations were performed mainly on the PC
cluster system of the Institute of Physics in Zagreb, donated by the Alexander
von Humboldt Foundation.

\end{acknowledgments}


\end{document}